\documentstyle[aps,prb]{revtex}

\begin{document}

\twocolumn[\hsize\textwidth\columnwidth\hsize\csname
@twocolumnfalse\endcsname

\draft
\author{Patricia R. Levstein$^{1}$, Gonzalo Usaj$^{1}$ and Horacio M. Pastawski$\thanks{corresponding author. E-mail: horacio@famaf.unc.edu.ar}^{1,2}$}
\address{$^{1}$Facultad de Matem\'{a}tica, Astronom\'{i}a y F\'{i}sica.Universidad
Nacional de C\'{o}rdoba. Ciudad Universitaria. 5000 C\'{o}rdoba- Argentina\\
$^{2}$International Center for Theoretical Physics, P.O. Box 586. 34100
Trieste, Italy}
\title{Attenuation of polarization echoes in NMR: A study of the emergence of
dynamical irreversibility in many-body quantum systems}
\maketitle

\begin{abstract}
The reversal of the time evolution of the local polarization in an
interacting spin system involves a sign change of the effective dipolar
Hamiltonian which refocuses the ``spin diffusion'' process generating a {\it polarization echo}. Here, the attenuation of these echo amplitudes as a
function of evolution time is presented for cymantrene and ferrocene {\it polycrystalline} samples, involving one and two five spin rings per molecule
respectively. We calculate the fraction of polarization which is not
refocused because only the secular part of the dipolar Hamiltonian is
inverted. The results indicate that, as long as the spin dynamics is
restricted to a single ring, the non-inverted part of the Hamiltonian is not
able by itself to explain the whole decay of the polarization echoes. A
cross over from exponential (cymantrene) to Gaussian (ferrocene) attenuation
is experimentally observed. This is attributed to an increase of the
relative importance of the spin dynamics, as compared with irreversible
interactions, which favors dynamical irreversibility.
\end{abstract}

] \narrowtext

\section{Introduction}

In a network of dipolar coupled identical spins in the solid state, the
flip-flop mechanism leads to a polarization dynamics often described as
``spin diffusion''. However, an ingenious NMR\ pulse sequence\cite{Rhim}
made evident that this many body evolution can be reversed by changing the
sign of the effective Hamiltonian (${\cal H}\rightarrow -\left[ 2\right] 
{\cal H}$). This practical realization of the Loschmidt daemon\cite{Rhim} is
possible because the sign and magnitude of the dipolar interaction depends
on the orientation of dipoles relative to the internuclear axis. In recent
years, it was shown\cite{ZME} that {\it local polarization} in a network of $%
^1$H spins can be monitored using the rare $^{13}$C as local probes. The
local excitation is allowed to ``diffuse'' during a time $t_R$ with ${\cal H}
$ and then it continues the evolution with $-\left[ 2\right] {\cal H}.$ The
polarization maxima appearing at around $t_R+\left[ \frac 12\right] t_R$ are
called {\it polarization echoes}. In practice, however, recovery of
polarization is not total. In fact, the echo amplitudes attenuate with
increasing $t_R$. Even when the origin for the decay of these echoes can not
be easily determined, when compared for example with Hahn echoes, it was
suggested \cite{ZME} that the characteristic attenuation of the echo
amplitudes might be useful for the study of molecular dynamics in solids.
Since then, the sequence for indirect measurement of proton spin diffusion
in the rotating frame has been applied to static and rotating samples (MAS)
as an attractive tool for structural characterizations of solids \cite
{zme2,hir,tomaselli}. It has also been used in its laboratory frame version
to observe quantum beats and interference phenomena in molecules with few
degrees of freedom \cite{PLU,PUL}. In this last case, refocusing by external
means was not the main issue, but a tool to compensate undesired evolution
in the proton system while the $^{13}$C is acting as a local probe. Thus,
the sequence succeeded in showing a spin dynamics which is a fingerprint of
the dipolar coupled system itself.

On the other hand, the pulse sequence to reverse the dipolar evolution of
spin polarization might be a powerful experimental tool to test how
dynamical irreversibility emerges in many-body quantum systems. This could
deepen our understanding of the mechanisms leading to the observed time
asymmetric behavior of macroscopic systems\cite{lebo}. From numerical
simulations, it is well known\cite{Izrailev} that {\it classical} systems
with chaotic behavior evolve showing ``diffusive'' properties. If at a time $%
t_{R}$ the motion is reversed, the trajectory can not be retraced but for a
short period of time. This strong limitation to practical reversibility
appears as a consequence of the {\it instability} which amplifies the
numerical round-off errors. Such amplification of instabilities, which can
be expected to grow with system size, explain why for {\it many-body} {\it %
classical }systems, the reversed evolution miss the initial macrostate \cite
{Bellemans} by an amount which increases with $t_{R}$, the time when
velocities are inverted. This contrasts with the apparent greater numerical
reversibility of some {\it simple quantum} models, where the motion appears
to recur to the initial state with an accuracy comparable with those errors.
Even more, the application, at the reversal time $t_{R}$, of quite large
random distortions of the relative phases of the states that expands the
evolved state, is not able to destroy this ``anti-diffusive'' evolution\cite
{Izrailev}. General solutions for{\it \ many-body quantum } systems become
much more difficult to obtain. Hence, there is an obvious interest in
experimental approaches to test possible hypothesis which might help to
handle those cases. It is from this point of view that several questions
arise: What is the functional dependence on $t_{R}$ of the polarization
echoes? Is the attenuation rate for the polarization echoes related to any
other independently measurable parameter? What kind of information can be
obtained from the functional form and degree of the attenuation? Does the
reversal of the spin dynamics show any instability with respect to natural
fluctuations characterizing the interaction with the environment? Can we
learn anything else about processes leading to irreversibility or ``true
relaxation'' in magnetic resonance by using this sequence? This paper must
be considered a first effort addressing these fundamental questions.

The sequence to reverse the dipolar evolution of a local excitation was
successfully applied \cite{ZME,zme2} to a single crystal of ferrocene \cite
{Seiler}, Fe(C$_5$H$_5$)$_2$. In this system the hierarchy of $^1$H dipolar
couplings was exhaustively analyzed \cite{PUL}, leading to the conclusion
that both the intra and inter-ring interactions within a molecule are
important, with a further increase in the connectivity of the dipolar
network given by the intermolecular interactions (see Fig. 1). A step to
simplify the topology of the coupling network while keeping the ring
structure as a building block, consists in the use of cymantrene \cite{Fiz},
Mn(C$_5$H$_5$)(CO)$_3.$ Having a single ring per molecule, it excludes an
important portion of the couplings (see Fig. 2). From a practical point of
view, if the pulse sequence is going to be useful to extract information in
a wide range of different dipolar coupled systems, it should be possible to
work with polycrystals. In this work, a special effort has been devoted to
get as much detail as possible from polycrystalline samples.

\section{Experimental Methods}

All the NMR measurements on polycrystalline samples of ferrocene, Fe (C$_5$H$%
_5$)$_2$, and cymantrene, (C$_5$H$_5$)Mn(CO)$_3$, with natural $^{13}$C
isotopic abundance were performed at room temperature. At this temperature,
the cyclopentadienil rings of both, ferrocene and cymantrene, perform fast
rotations around their five-fold symmetry axis with a very short correlation
time ($\approx 10^{-12}s$ for ferrocene) \cite{fast} giving rise to axially
symmetric $^{13}$C-NMR spectra as shown in Fig. 3.

We used a Bruker MSL-300 spectrometer, equipped with a standard Bruker
CP-MAS probe, operating at a $^{13}$C frequency of approximately 75.47 $MHz$%
. This frequency was finely adjusted in each system to correspond to the
resonance of molecules with their five-fold molecular axis perpendicular to
the external magnetic field, as indicated by an arrow in Fig. 3. In this
situation, the most intense point of the axially symmetric spectrum, shown
by both systems, ($\sigma _{\perp }$) is exactly on resonance simplifying
data analysis. We used the pulse sequence proposed by Zhang et al.\cite{ZME}
for refocusing spin polarization sketched in Fig. 4. After cross
polarization from the abundant $^1$H spins ($I$-spins) to the rare $^{13}$C (%
$S$-spin) during the time $t_c$, the $S$ magnetization is kept spin locked
for a time $t_s$ during which, as we will see later, the $I$-spin coherence
decays to less than 2$\%$ of its initial value. During time $t_d$ cross
polarization from $S$ to $I$ should polarize mainly the nearest neighbor $I$
spin (i.e. the $^1$H directly bonded to a $^{13}$C). This time was
accurately determined to be 87 $\mu s$ and 95 $\mu s$ for the $\sigma
_{\perp }$ resonance frequency in Fe (C$_5$H$_5$)$_2$ and (C$_5$H$_5$)Mn(CO)$%
_3$, respectively, as the time at which the polarization transfer reaches
its first maximum in a simple cross polarization sequence with variable
contact time. The experimental results of these measurements are displayed
in Fig. 5 and 6.

Thus, after the time $t_d,$ only those protons directly bonded to a $^{13}$C
can contribute to the spin polarization. This polarization is locked during
a time $t_1$ at which dipolar evolution occurs in the rotating frame\cite
{Levitt} with a Hamiltonian : 
\begin{equation}
{\cal H}_{II}^{^{\prime }}=-\left[ \frac 12\right]
\sum_{j>k}\sum_k^{\,\,}d_{jk}\,\left[ 2I_j^{\,\,y}I_k^{\,\,y}-\frac 12\left(
I_j^{+}I_k^{-}+I_j^{-}I_k^{+}\,\right) \right] .  \label{Hii'}
\end{equation}
Here the $I_j^{+(-)}$ operators are quantized referring to the $y$ basis 
\cite{PLU}, and the interaction parameters are 
\begin{equation}
d_{jk}^{}=-\frac{\mu _{0\,\,}\gamma _I^2\,\,\,\,\hbar ^2}{4\pi r_{jk}^3}\,\,%
\frac 12\left\langle 3\cos ^2\theta _{jk}-1\right\rangle ,  \label{d}
\end{equation}
where $\gamma _I^{}$ is the gyromagnetic factor of the $I$-spin, the $%
r_{jk}^{}$'s are their internuclear distances and $\theta _{jk\text{ }}$are
the angles between internuclear vectors and the static magnetic field. As
mentioned above, a simplifying fact of these systems is that the rings
perform fast rotations around the five-fold symmetry axis and therefore the
interaction parameter is time averaged \cite{fast}. The new constants depend 
\cite{Slichter} only on the angle $\theta $ between the molecular axis and
the external magnetic field $B_0$ and the angles $\gamma _{jk}$ between the
internuclear vectors and the rotating axis. A following ($\pi /2$)$_x$ pulse
leads the system back to the laboratory frame where evolution in the reverse
sense occurs during the time period $t_2$ with the Hamiltonian:

\begin{equation}  \label{Hii}
{\cal H}_{II}=\sum_{j>k}\sum_k^{\,\,}d_{jk}\,\left[ 2I_j^{\,\,z}I_k^{\,\,z}-%
\frac 12\left( I_j^{+}I_k^{-}+I_j^{-}I_k^{+}\,\right) \right] .
\end{equation}

A ($\pi /2$)$_{-x}$ pulse followed by another short cross-polarization time $%
t_{p\text{ }}$produces again a spatially selective polarization transfer
from $I$ to $S$ spins. The $S$-spin FID is then acquired under high
resolution conditions, i.e. by keeping a spin lock on the $I$-spin system.

\section{Results and Discussion}

As a first step in our study we have analyzed the evolution of $I$
polarization in a polycrystalline sample of ferrocene as an irreversible
spin diffusion, following the main ideas in the pioneering work by
M\"{u}ller et al.\cite{Muller}. In that work, the cross polarization in a
single crystal of ferrocene was described by the Hamiltonian:

\begin{equation}
{\cal H}=\hbar \left[ \Omega _II_{}^z-\gamma _IB_{1I}^yI_{}^y+\Omega
_SS_{}^z-\gamma _SB_{1S}^yS_{}^y\right] +b2I_{}^zS_{}^z,  \label{Hcp}
\end{equation}
where $\Omega _j$ are the resonance offsets, $B_{1I}^y$ and $B_{1S}^y$ are
the RF field strengths, and

\begin{equation}
b=\frac{\mu _{0\,\,}\gamma _I\,\gamma _S\,\hbar ^2}{4\pi r_{IS}^3}\frac 12%
\left\langle 3\cos ^2\theta -1\right\rangle .  \label{b}
\end{equation}
This Hamiltonian considers only the dipolar interactions of the isolated $S$
spins with the directly bound $I$ spins. No dipolar interactions between $I$
spins are considered in it. They are taken into account, in a
phenomenological way, through an {\it isotropic} spin-diffusion process with
rate $R$ leading to a magnetization that, for $\Omega _j=0$ and under exact
Hartmann-Hahn \cite{HH}condition $\gamma _IB_{1I}^y$=$\gamma _SB_{1S}^y$, is
expressed as:

\begin{equation}
M_S^y(t)=A\left\{ 1-\frac 12\exp (-Rt)-\frac 12\exp (-3Rt/2)\cos (bt/\hbar
)\right\} ,  \label{msx}
\end{equation}
where $A$ depends only on temperature and Zeeman splitting.

Thus, the oscillations with frequency $\omega =b/\hbar ,$ experimentally
observed were successfully explained. In our study, we can select molecules
with their rotational axes at some specific angle $\theta $ with respect to
the external magnetic field $B_0$ and perform the fitting for each
orientation, although strictly speaking only $\theta =90^{\circ \text{ }}$
is exactly {\it on resonance}. An improvement in these fittings can be
performed by considering that ``spin diffusion'' is not a consequence of
fluctuations of isotropic fields but fields that, being essentially {\it %
dipolar} in origin, have the precise anisotropy given by Eq. (\ref{Hii}).
This leads to the new expression: 
\begin{equation}
M_S^y(t)=A\left\{ 1-\frac 12\exp (-Rt)-\frac 12\exp (-2Rt)\cos (bt/\hbar
)\right\} .  \label{Msxa}
\end{equation}
Figure 5 shows the experimental points and calculated curves (solid lines)
for molecules with $\theta =\frac \pi 2$ (upper set of data) and with the
``magic'' angle $\theta _{m\text{ }}$ $=\arccos 1/\sqrt{3}$ (lower set).
Note that no extra free parameters are added in Eq. (\ref{Msxa}). It can be
seen that the experimental curve does not decrease, especially in the first
valley, as much as predicted by Eq. (\ref{msx}) (dotted line) or its dipolar
version Eq. (\ref{Msxa}).

Further improvement of the fitting, though at high computational cost, has
been reached by incorporation of an exact calculation of the intra-ring $^1$%
H-spin dynamics. The better fitting (dashed line in Fig. 5) is explained
because the magnetization ``hole'' produced in the $I$ polarization because
of the transfer to its directly bound $S$ , is partially ``filled'' by
magnetization from other protons in the ring during that short time. A new $%
R $ parameter accounts for fluctuating fields due to inter-ring interactions.

We have performed non-linear least square fittings of the experimental
points to Eq. (\ref{Msxa}) for the whole frequency spectra. That is,
fittings equivalent to those shown in Fig.5 with solid lines, were performed
in frequency steps of $\approx 80Hz$ covering the whole $^{13}$C spectrum of
ferrocene for contact times ranging from 2$\mu s$ to 3$ms$. The $A,R$ and $%
b/(2\pi \hbar )$ parameters obtained from these fits are shown in Fig.7. The
first observation giving reliability to the results is that $A$ reproduces
the spectral line-shape obtained at long cross-polarization contact times.
This is shown in the upper panel of Fig. (7) where the spectrum obtained
with 2$ms$ contact time is superimposed to the $A$ values. The second fact
is that $b$ shows the expected ($3\cos ^2\theta -1$) dependence (except in
the region where $b/\hbar \approx 0\ll R$, where the assumptions of the
model break down), which is the same as that of the resonance frequencies.
This is emphasized in the graph by tracing the resulting linear dependence
between $b$ and frequency.

There is no doubt that $R$ is a parameter giving information on the $I$-$I$
dipolar coupling network. The shape of $R$ vs. frequency does not change
much by considering it as an isotropic relaxation parameter or an
anisotropic (dipolar in origin) one. Its magnitude, however, does change.
The dependence of $R$ on frequency constitutes an evidence that, in contrary
to the suggestions in Refs. 2 and 3, the ``$^1$H-spin diffusion'' is not
restricted to one ring, not even to the complete molecule. In such a simple
case, we would have a single averaged dipolar coupling proportional to ($%
3\cos ^2\theta -1$) as is the case for the $I$-$S$ dipolar coupling $b$ (see
lower panel of Fig.7). Besides, $R$ is a useful parameter in order to
distinguish intra and inter-molecular contributions to the ``spin
diffusion''. For example, at the frequency corresponding to molecules with
their axes at the magic angle with respect to $B_0$, $R$ is entirely due to
intermolecular couplings. Being more specific, only the magnetically
non-equivalent molecules contribute, because even when there is
communication with the equivalent ones the spin-diffusion in them is
cancelled out ($R_{{\rm intra}}\approx 0$). As can be observed in the
central panel of Fig. 7, the $R$ value at $\theta _m$ is approximately one
eighth of the maximum value. This maximum does not appear when the
intramolecular coupling maximizes, but when the balance between both (intra
and inter-molecular) contributions becomes maximum. Besides, it is
worthwhile to note that working on a polycrystalline sample, each $R(\theta
) $ value involves an average of molecular orientations over a cone of angle 
$\theta ,$ i.e. having the same angle with respect to $B_0$. In terms of
characteristic times, the maximum $R^{-1}\approx 2.7ms$ is at $\theta _m$
while the minimum $R^{-1}\approx 0.37ms$ occurs at $\theta \approx 48^{\circ
}.$ At the on-resonance frequency, corresponding to $\sigma _{\bot }$ in our
experiments, $R^{-1}\approx 0.73ms$. Although signal to noise ratio for
cymantrene is far from the ideal situation offered by ferrocene, an
acceptable fitting is still achieved at $\sigma _{\bot }\,$ giving $%
R^{-1}>1.2ms.$ The comparison of this value with that for the same
orientation in ferrocene manifests the less dense network of $I$-spins in
cymantrene.

The local $I$-polarization in ferrocene as detected in the $S$-spin by using
the pulse sequence sketched in Fig. 4 is displayed in Fig. 8. Experimental
parameters are given in the caption. It can be noticed that for each delay
time $t_1$ a maximum in the local polarization, called polarization echo, is
obtained at a time $t_2\approx \left[ \frac 12\right] t_1+t_m$, where $t_m$ $%
\approx \left[ \frac 12\right] (t_p+t_d)/2$ is the time spent to compensate
for the $I$-spin evolution occurring during the short cross polarization
pulses $t_d$ and $t_p$. It is not difficult to see that at $t_1\approx 240$ $%
\mu s$ almost 50\% of the initial polarization has not been able to refocus,
reflecting the irreversible part of the evolution which leads to the decay
of the polarization echo amplitudes. Superimposed to the curves, there are
some high-frequency ($\approx 59kHz$) oscillations which are particularly
noticeable around the maximum amplitude of the echo. These oscillations are
well above the experimental error and its quantum origin has been explained
previously \cite{PUL}. A similar study was performed in cymantrene, where
the curve for $t_1=0$ is displayed in Fig.9. It can be seen there, that the
quantum interferences are better developed than in ferrocene: a quantum beat
at $400$ $\mu s$ reflects the discrete nature of the system where the $I$%
-spin evolution occurs, while a mesoscopic beat insinuated at $600$ $\mu s$
indicates the finite size of the ring. This can be understood in terms of
the better isolation of one cyclopentadienil ring protons from other ring
protons, as can be inferred from the crystalline structure of cymantrene
(Fig. 2) as compared with ferrocene (Fig. 1). Besides, recording of the
local polarization around their maxima in cymantrene, for the same delay
times $t_1$ studied in ferrocene, allows us a comparative analysis of the
attenuation of polarization echoes. The experimental amplitudes of the
polarization echoes $M_{PE}$ obtained for each system are plotted in Fig. 10
as a function of the total forward evolution $t_R.$ According to the
discussion above we have $t_R=\frac 23(t_1+t_2^M+2t_m)\approx t_1+2t_m$ with 
$t_1=0\mu s$, $\,80\mu s$,\thinspace \thinspace $160\mu s$ and $240\mu s$
and $t_2^M$ the time at which the maximum polarization occurs. The
experimental data clearly show different functional dependence of the decay
of the echo amplitude on $t_R$. Data of Ferrocene are Gaussian-like while
those of Cymantrene are exponential. According to our previous work\cite
{PLU,PUL} the correct normalization of the data requires $%
M_{PE}(t_R=0)\equiv 1$. This allows us to compare different systems. It is
well known that decoherence would lead to an ergodic distribution of
polarization\cite{BruschErnst}. In Ferrocene, the inter-ring and
inter-molecular $I$-$I$ interactions are both very important\cite{PUL}, then
one expects a significant spreading of the refocused polarization. By
choosing an asymptotic value of 0 a good fittings to a Gaussian is obtained
with a characteristic time as the only free parameter. In contrast, the
rings of cymantrene are better isolated within the experimental times $%
t_R\leq 335\mu s$ . Then we choose the asymptotic value of 1/5 for the
exponential decay obtaining a good fitting.

Irreversible interactions are those we have not been able to reverse in our
experiments. This is the situation of the non-secular terms in the
Hamiltonian. In order to study whether they are responsible for the
attenuation of the polarization echoes, we performed exact calculations of
the spin dynamics using parts A to F of the sequence sketched in Fig.4 for a
cyclopentadienil ring with a single $^{13}$C. While the $\pi /2$ pulses are
considered ideal, evolution during the short cross polarization times $t_{d%
\text{ }}$ and $t_p$ is described with the complete Hamiltonian following
the scheme by Levitt et. al. \cite{Levitt}. During time $t_1$ the
Hamiltonian is:

\begin{equation}
{\cal H}_{}^1={\cal H}_I+{\cal H}_{II}^{\prime }+{\cal H}_{II}^{\prime
\prime }+{\cal H}_{IS},  \label{ht1}
\end{equation}
where 
\begin{equation}
{\cal H}_I=-\hbar \gamma _IB_{1I}^y\sum_kI_k^y  \label{hi}
\end{equation}
and 
\begin{equation}
{\cal H}_{IS}=\sum_kb_k2I_k^zS_{}^z.  \label{his}
\end{equation}
The contribution ${\cal H}_{II}^{\prime \prime }$ takes into account the
non-secular part of the $I$-spin dipolar interactions which were neglected
in Eq. (\ref{Hii'}). On the other hand, evolution during time $t_2$, taking
into account both $\pi /2$ pulses on the $I$-system, can be described by: 
\begin{eqnarray}
{\cal H}_{}^2 &=&\overline{X}(\frac \pi 2)\left\{ {\cal H}_{II}+{\cal H}_S+%
{\cal H}_{IS}\right\} X(\frac \pi 2)  \nonumber \\
&=&[-2]{\cal H}_{II}^{\prime }+{\cal H}_S+\sum_kb_k2I_k^yS_{}^z,  \label{ht2}
\end{eqnarray}
where ${\cal H}_{II}^{\prime }$ is given in Eq. (\ref{Hii'}) and

\begin{equation}
{\cal H}_{S}=-\hbar \gamma _{S}B_{1S}^{y}S_{{}}^{y}.  \label{hs}
\end{equation}
Fig. 11 shows the results of the complete sequence calculation. The
attenuation generated is much slower than the experimentally observed,
giving a decay of less than 10\% for times where the experiment shows around
50\%. It was verified by excluding one term at a time that the only term
producing a visible attenuation of the polarization echoes is ${\cal H}_{IS}$
during $t_{1}.$ This term transforms some $I$-spins polarization into $I$-$S$
coherence. Increase on the size of the system to six, seven, and eight
protons shows no faster decays than the ring with five protons, as long as
the ring topology is kept and a single $S$ is included in the spin dynamics.
If those non-invertible terms were responsible for the attenuation of the
polarization echoes in ferrocene, the spin dynamics in the actual complex
proton network should play a relevant role amplifying their effects. Due to
the magnitude of the computation involved, we have not been able so far to
treat other topologies of the $I$-network in a systematic way.

\section{Conclusions}

We used the structural properties of cymantrene and ferrocene, which
determine their different networks of dipolar coupled hydrogens, to test how
this topology influences the ``diffusion'' of a local excitation. We also
studied the polarization echoes which describe the ``anti-diffusion''
process or backward evolution. With respect to the evolution forward in time
, we have seen that cymantrene gives better developed quantum beats when
compared with ferrocene. This is caused by the fact that the $I$-$I$
couplings network is better limited to one ring. However, from the
attenuation of the oscillations we conclude that there is still an
appreciable amount of $I$-$I$ intermolecular couplings acting at
intermediate times. With regard to the attenuation of the polarization
echoes, cymantrene presents an {\it exponential} decay in contrast with
ferrocene where a {\it Gaussian } decay of the polarization echoes is
observed.

The exponential regime is quite clear when there is an irreversible process
with a characteristic time $\tau _\phi $ shorter than the typical time for
the spin dynamics. Then, the probability to return to the initial state
would be given by the survival probability, which is proportional to $\exp
[-t/\tau _\phi ].$ This seems to be the case for cymantrene, where a{\it \
strong }local source of relaxation could be traced back to the quadrupolar
nature of the Mn nucleus (nuclear spin 5/2).

The interpretation of the Gaussian decay of the polarization echoes observed
in ferrocene is more subtle. The absence of obvious irreversible
interactions requires the consideration of small non-inverted terms, the
most significative of these are the $I$-$S$ non-secular terms. However, as
long as we consider a spin dynamics restricted to a single ring, those terms
are not enough to explain the decay which occurs in a time scale typical of
the dipolar dynamics. The ingredient that was left out from the calculation
is the complex topology of the actual network where the spin dynamics
occurs. This is expected to be very relevant at a time $t_{mb}$ at which the
quantum beats attenuate. In general, {\it weak }non-invertible terms produce
fluctuations at a rate $1/\tau _\phi $ leading to local spin flips whose
effect can be considered a damage or disease that propagates through the
network, due to the natural spin dynamics, like an epidemy. Thus, the damage
spreads with the ``spin diffusion'' rate. When the dipolar evolution is
reverted (by external means) the modified spin configurations can not
warrant a return to the initial state. That is, some reversible part of the
dynamics (e.g. the inter-ring dipolar interaction) paves the way to the
extraordinary efficiency of small irreversible fluctuations. This is
consistent with the results obtained by Lacelle\cite{Lacelle} for the
reversibility of classical spins with a dynamics described by a cellular
automaton model. The interplay between reversible dynamics and irreversible
interactions, resembles\cite{Keldysh} the process that causes electrical
resistivity in an impure metal at low temperatures. In this case, reversible
collisions with impurities (with rate $1/\tau _{{\rm imp}}$) favor the
effectiveness of the irreversible electron-phonon interaction (with rate 1/$%
\tau _\phi \ll $1/$\tau _{{\rm imp}}$) and set the diffusion coefficient $%
D\propto \tau _{{\rm imp}}$. Another analogous situation is the unstable
quantum oscillator \cite{Paz} where the reversible potential favors
dissipation. Since the study of dynamical irreversibility \cite{lebo} in
many body quantum systems is still in its infancy, there is much room for
numerical and analytical studies to verify the main suggestion of our
experiment: that these systems can present a local instability amplifying
the natural fluctuations of the environment.

It is interesting to emphasize that the systems experimentally studied show
a crossover between the different physical regimes while the attenuation of
the polarization echo occurs on the same time scale. This is possible
because the slower spin dynamics of cymantrene is compensated by its much
stronger fluctuating environment.

In summary we have shown the suitability of this new NMR pulse sequence \cite
{ZME} as a tool in the study of fundamental questions about dynamical
irreversibility. Our work has settled a conceptual framework allowing for
the identification of the relevant parameters governing such phenomenon,
that should stimulate the design of new experiments.

\section{Acknowledgments}

The authors express their gratitude to Professors R. R. Ernst, B. Meier, R.
Harris, for several stimulating comments as well as to M. Ernst , J. P. Paz
and J. Vollmer for fruitful discussions. HMP acknowledges F. Izrailev for
tutorial talks and H. Cerdeira for hospitality at ICTP.

This work was performed at the LANAIS de RMN (CONICET-UNC) with financial
support from Fundaci\'{o}n Antorchas, CONICOR, SeCyT-UNC and CONICET.

\section{Figure Captions}

{\bf Figure 1}: Crystalline structure of ferrocene, Fe(C$_5$H$_5$)$_2$ in
its (room temperature) monoclinic form with space group $P2_1/a$ as
described in ref. 11$.$ Two unit cells are included for comparison with Fig.
2.

{\bf Figure 2}: Crystalline structure of cymantrene (C$_5$H$_5$)Mn(CO)$_3$
according to crystallographic data on ref. 12$.$ The compound crystallizes
in the monoclinic space group $P2_1/a$ with four molecules per unit cell.

{\bf Figure 3: }$^{13}$C-NMR spectra of polycrystalline ferrocene after a
simple cross-polarization experiment with 85 $\mu s$ and 3$ms$ contact times
respectively. The arrow indicates the exact irradiation frequency. Note that
both spectra have essentially the same amplitude at $\sigma _{\perp }$ but
at short contact time there is almost no signal from molecules with their
rotational axis at the magic angle with the external magnetic field $B_0$.

{\bf Figure 4: }Pulse sequence for refocusing the evolution of $^1$H spin ($%
I $-spin) polarization. $S$-spin ($^{13}$C spin) polarization in the y-axis
is prepared by $I$-$S$ cross polarization. The $S$ magnetization is kept
spin locked for a time $t_s$ during which the $I$-spin coherence decays to
zero. A) Proton polarization is locally injected by cross-polarization from $%
S$ to $I$ during $t_d$. B) Dipolar evolution of the spin locked $I$-spin
polarization is allowed for a time $t_1$. C) With a ($\pi /2$)$_x$ pulse,
the proton magnetization is turned into the $z$-axis where D) dipolar
evolution occurs in the reverse sense during a period $t_2$. E) A ($\pi /2$)$%
_{-x}$ pulse, followed by F) a short cross-polarization time $t_p$ to the $x$%
-axis of the $S$-spins, allows for detection minimizing quantum
interferences between dashed and dotted pathways of {\it probability
amplitudes} of polarization. G) Acquisition is done under high resolution
conditions.

{\bf Figure 5: }$^{13}$C magnetization as function of contact time in a
simple cross polarization experiment in ferrocene powder. The amplitudes
have been frequency selected from the axially symmetric spectra. Data in the
upper part of the plot correspond to molecules with the five-fold axis
perpendicular to the external magnetic field ($\theta =\pi /2$), and the
lower part to molecules at the magic angle $\theta =\theta _m$. The
experiment was performed with $\omega _1/2\pi $=44.6$kHz$. The dotted curve
corresponds to a non-linear least square fitting to Eq. (\ref{msx}). Solid
lines are fittings to Eq. (\ref{Msxa}). Dashed line includes exact dynamics
in a ring besides the dipolar relaxation parameter.

{\bf Figure 6: }$^{13}$C magnetization as function of contact time in a
simple cross polarization experiment for cymantrene molecules with their
five-fold axes approximately perpendicular to the external magnetic field.

{\bf Figure 7 : }Parameters obtained by fitting all the experimental curves
of the type displayed in Fig.5 to Eq.(\ref{Msxa}). Fittings for orientations
differing in approximately $80Hz$ in the frequency spectra of ferrocene were
performed.

{\bf Figure 8:} Ferrocene experimental data as function of $t_1+$ $t_2$
using the sequence in Fig.4 with $\omega _{1I}/2\pi $=$44.6kHz$, $t_c=2ms$, $%
t_s$=$1ms$, $t_d=t_p=85\mu s$, for $t_1=0\mu s$, $80\mu s$, $160\mu s$ and
240$\mu $s.

{\bf Figure 9: }Cymantrene experimental data as function of $t_1+$ $t_2$
using the sequence in Fig.4 with $\omega _{1I}/2\pi $=$44.6kHz$, $t_c=2ms$, $%
t_s$=$1ms$, $t_d=t_p=95\mu s$, for $t_1=0\mu s.$

{\bf Figure 10: }Efficiency of the Loschmidt daemon. The properly normalized
experimental polarization echo amplitudes (maxima of the curves in Figs. 8
and 9) $M_{PE}$ for ferrocene and cymantrene as function of the time $t_R.$
Solid lines represent fittings of the experimental points to Gaussian and
exponential laws respectively.

{\bf Figure 11}: Theoretical calculation (to be compared with Fig. 8) for
the $^{13}$C polarization in a five proton ring of ferrocene after the pulse
sequence A-F sketched in Fig.4. The high frequency oscillations are
explained by the non-ideality of the $^{13}$C probe and they have an
approximate frequency of $\omega _{1I\text{ }}$ . They are quantum
interferences caused by the failure of full transfer of polarization between 
$I$ and $S$ spins.

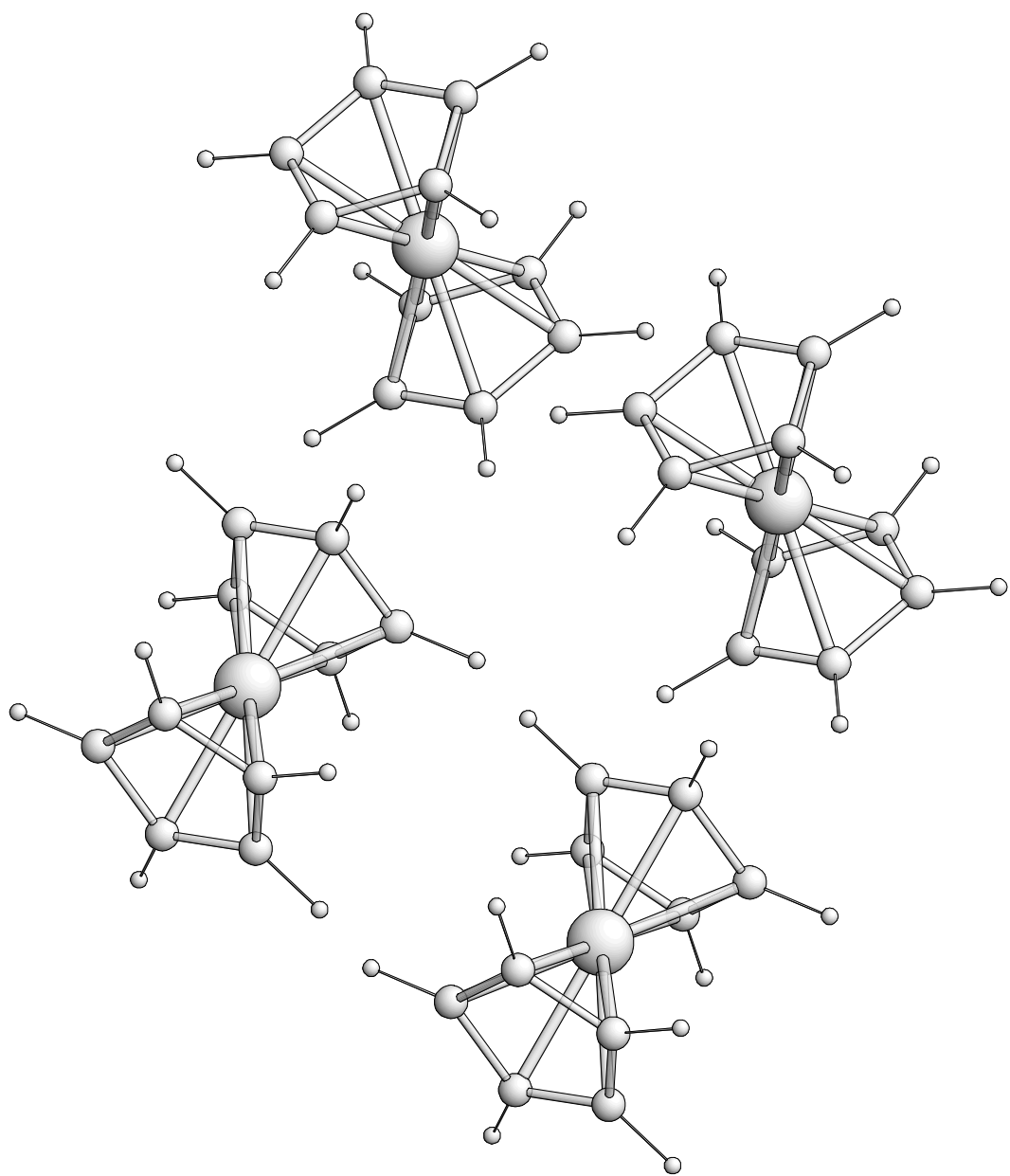
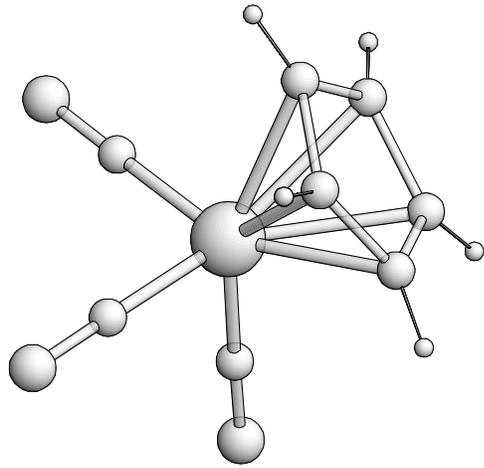
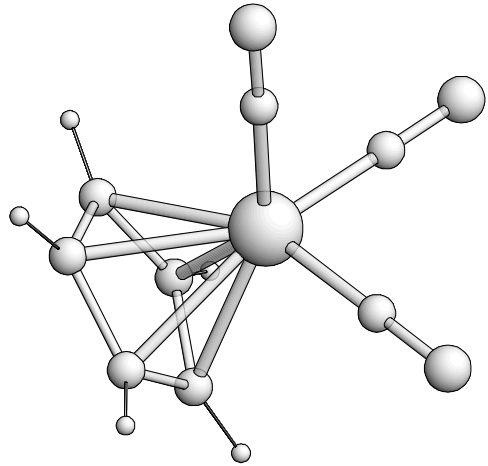
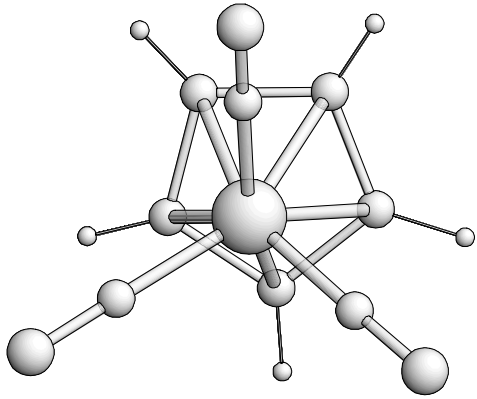
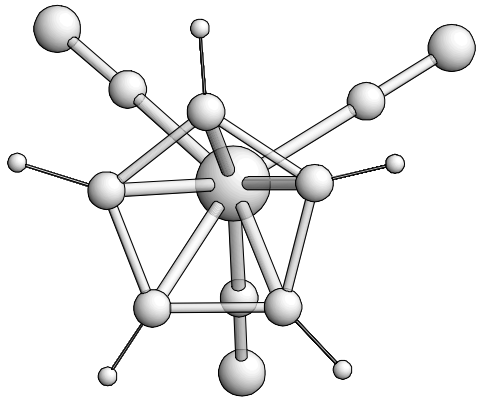
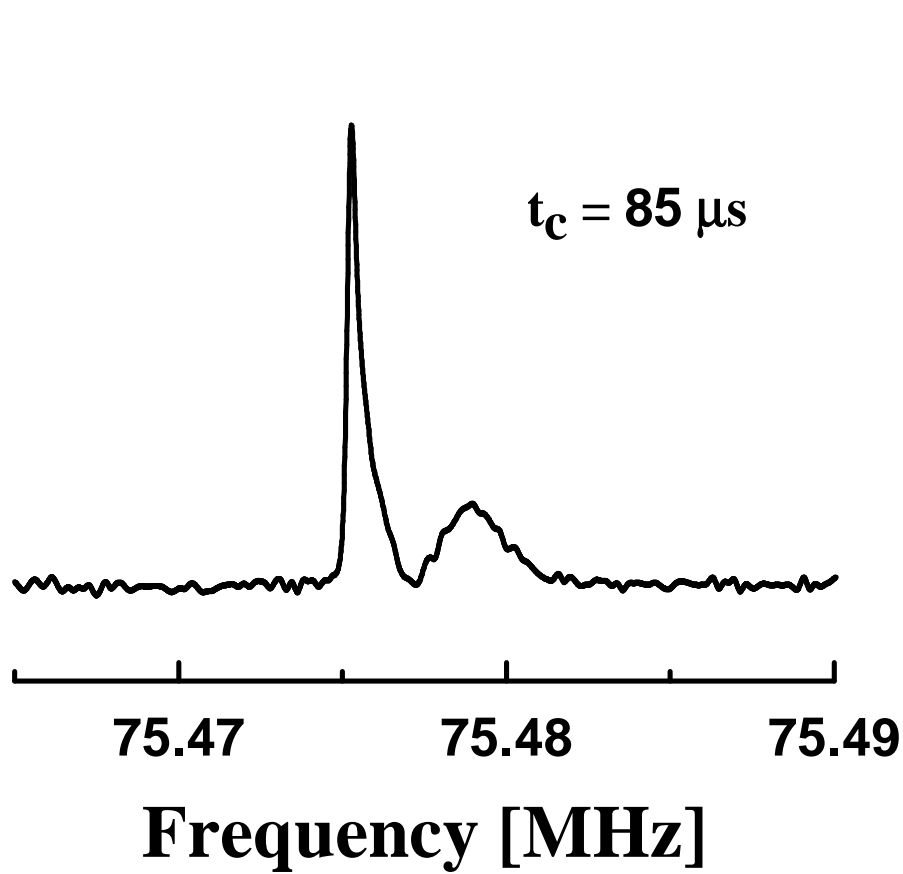
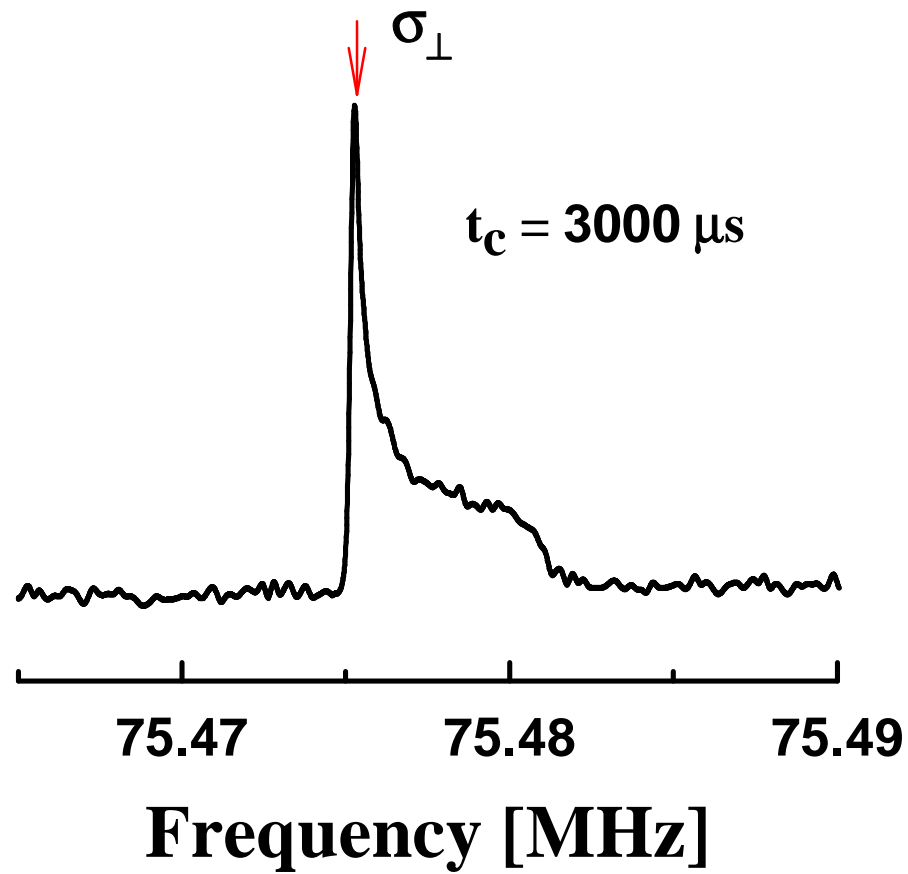
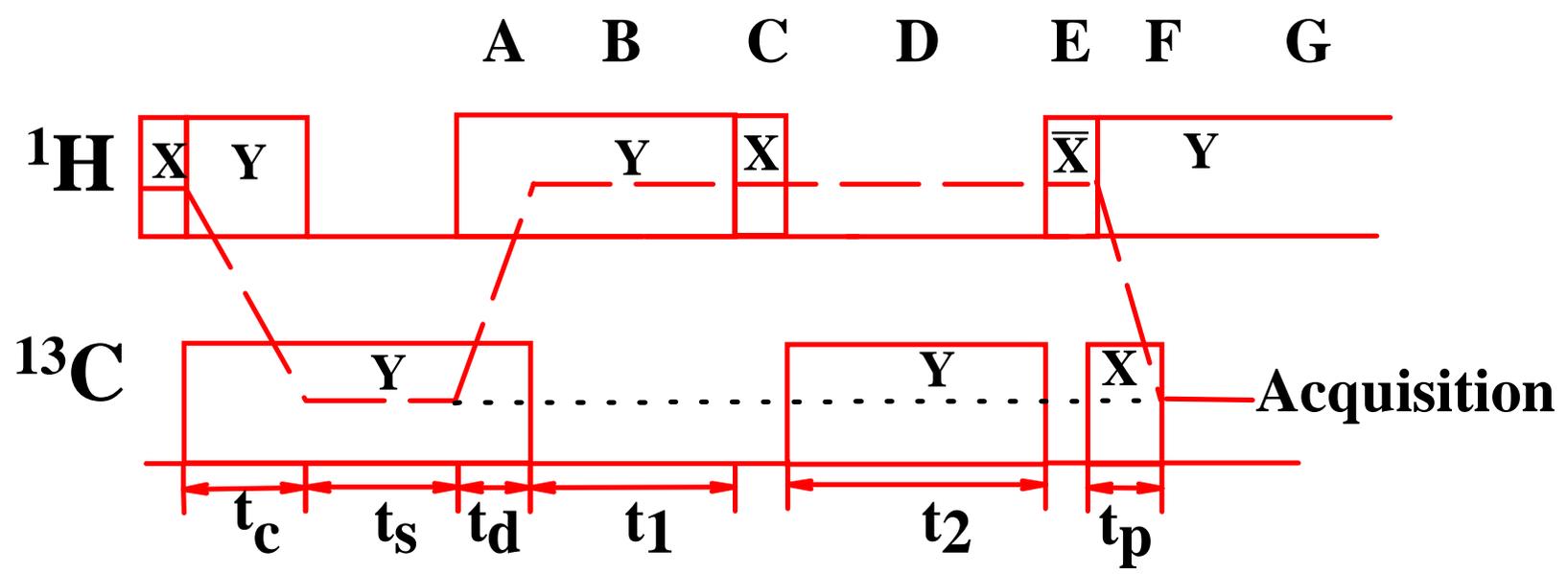
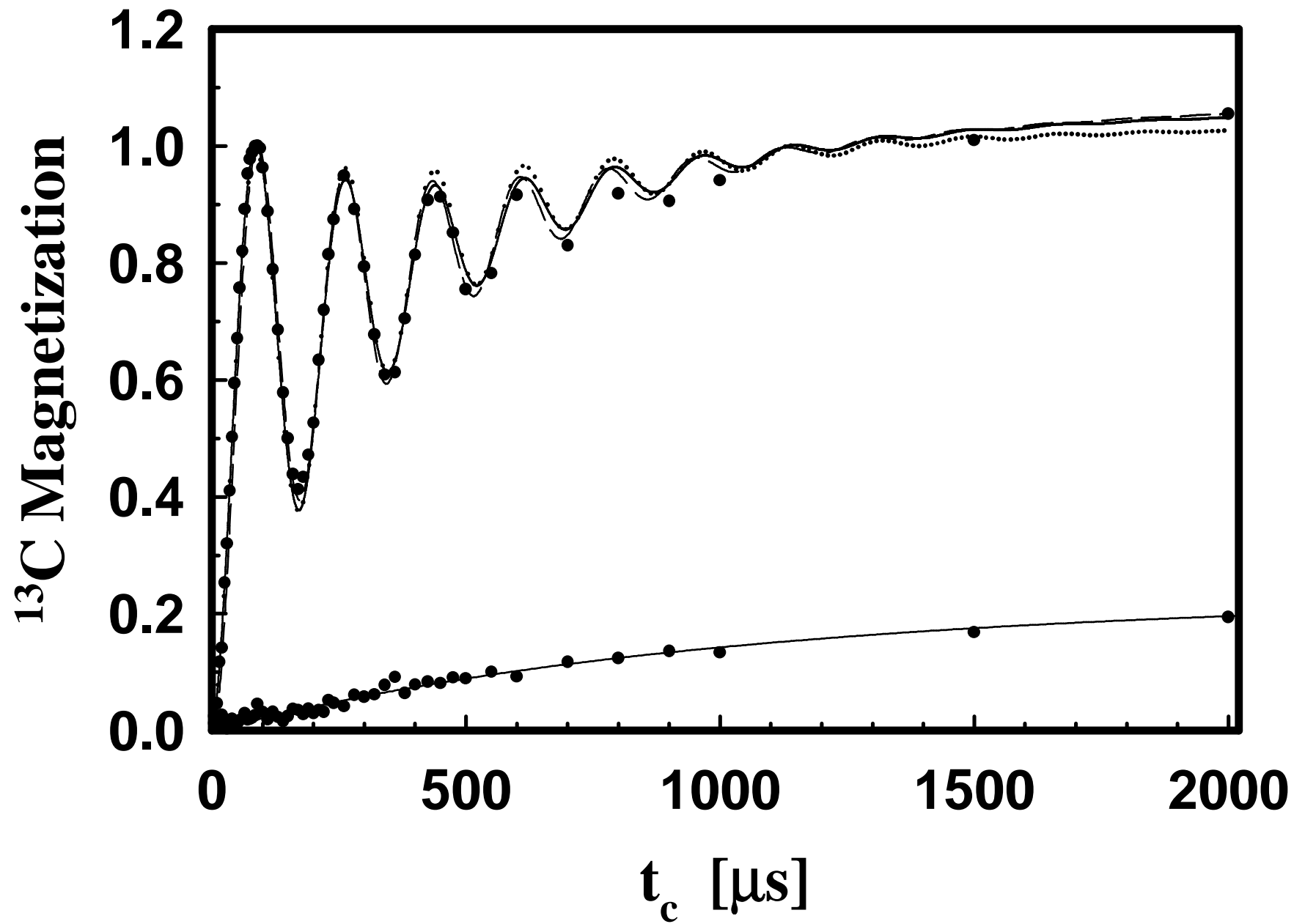
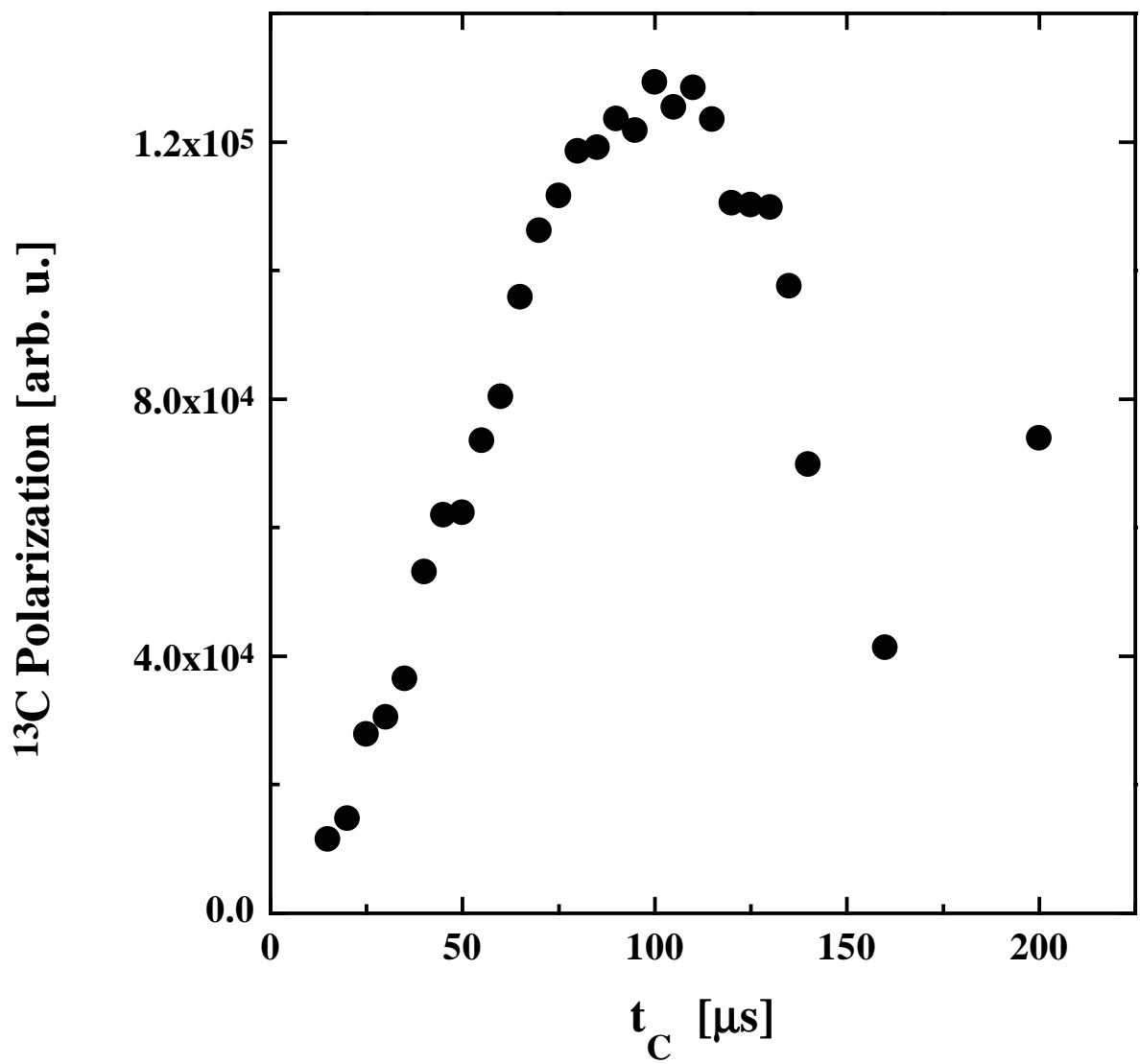
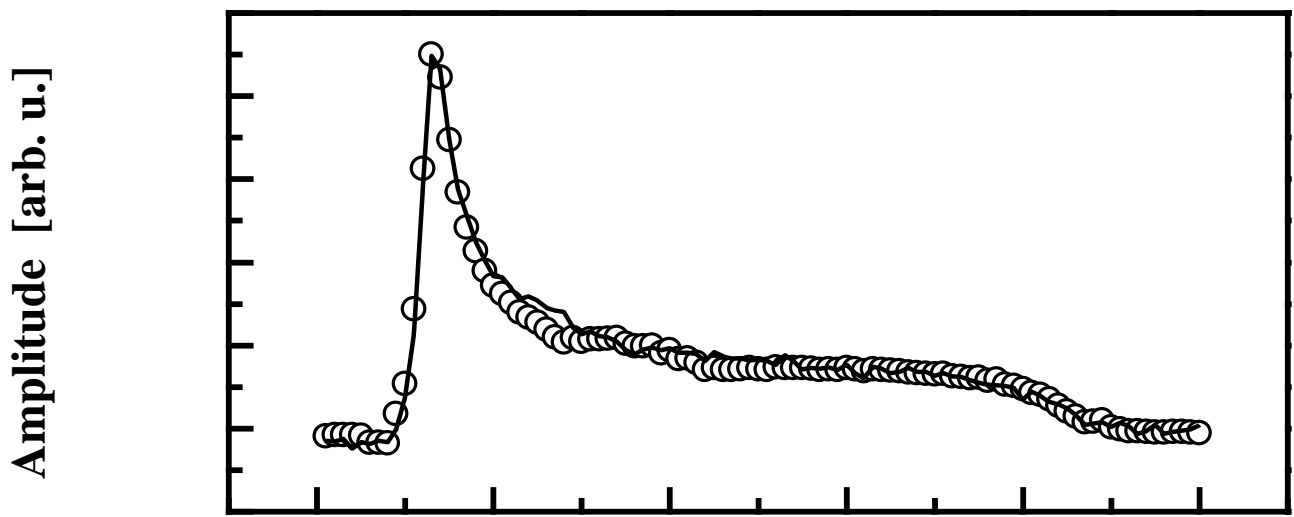
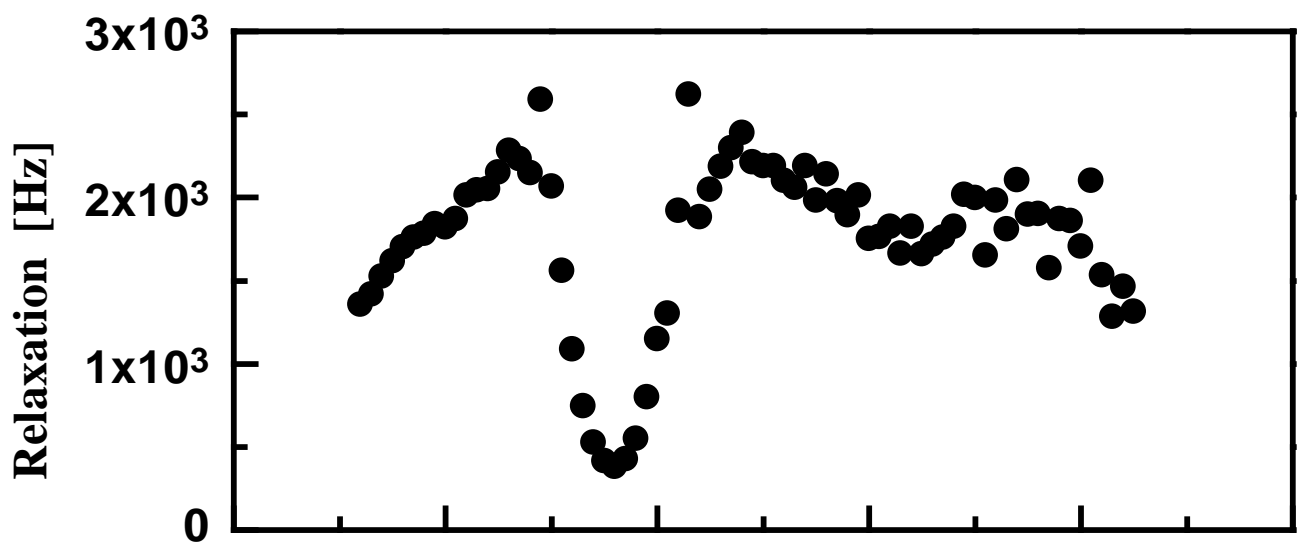
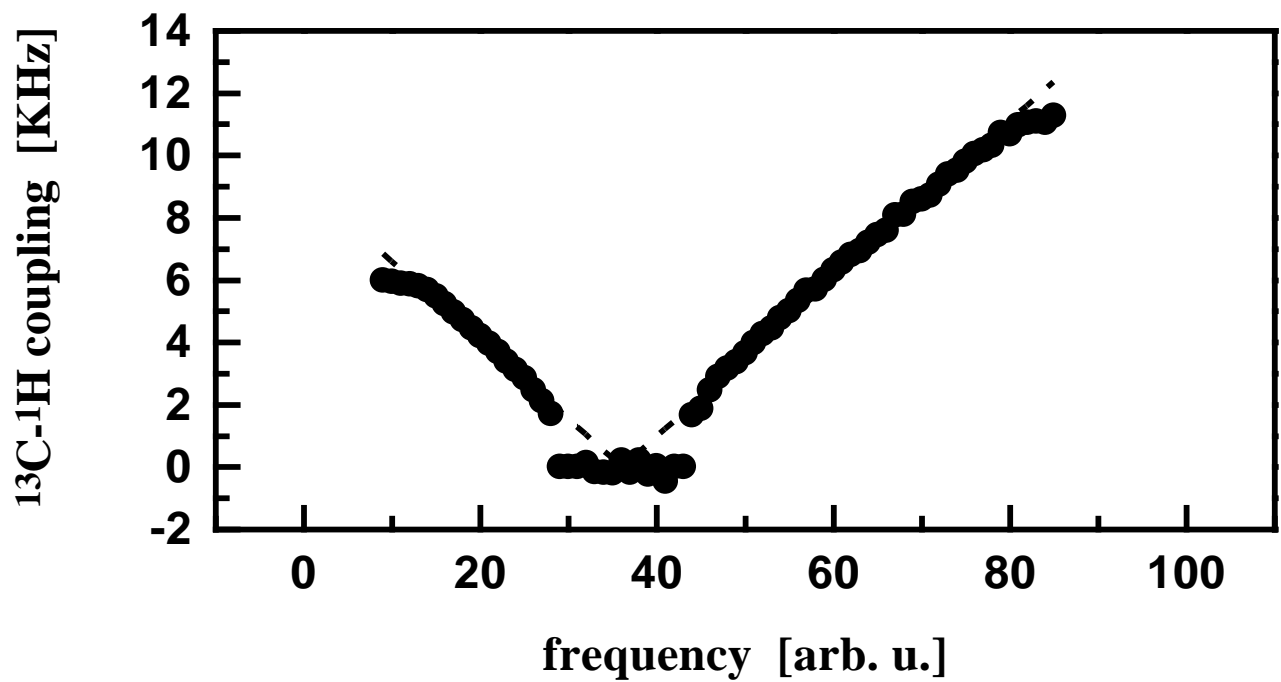
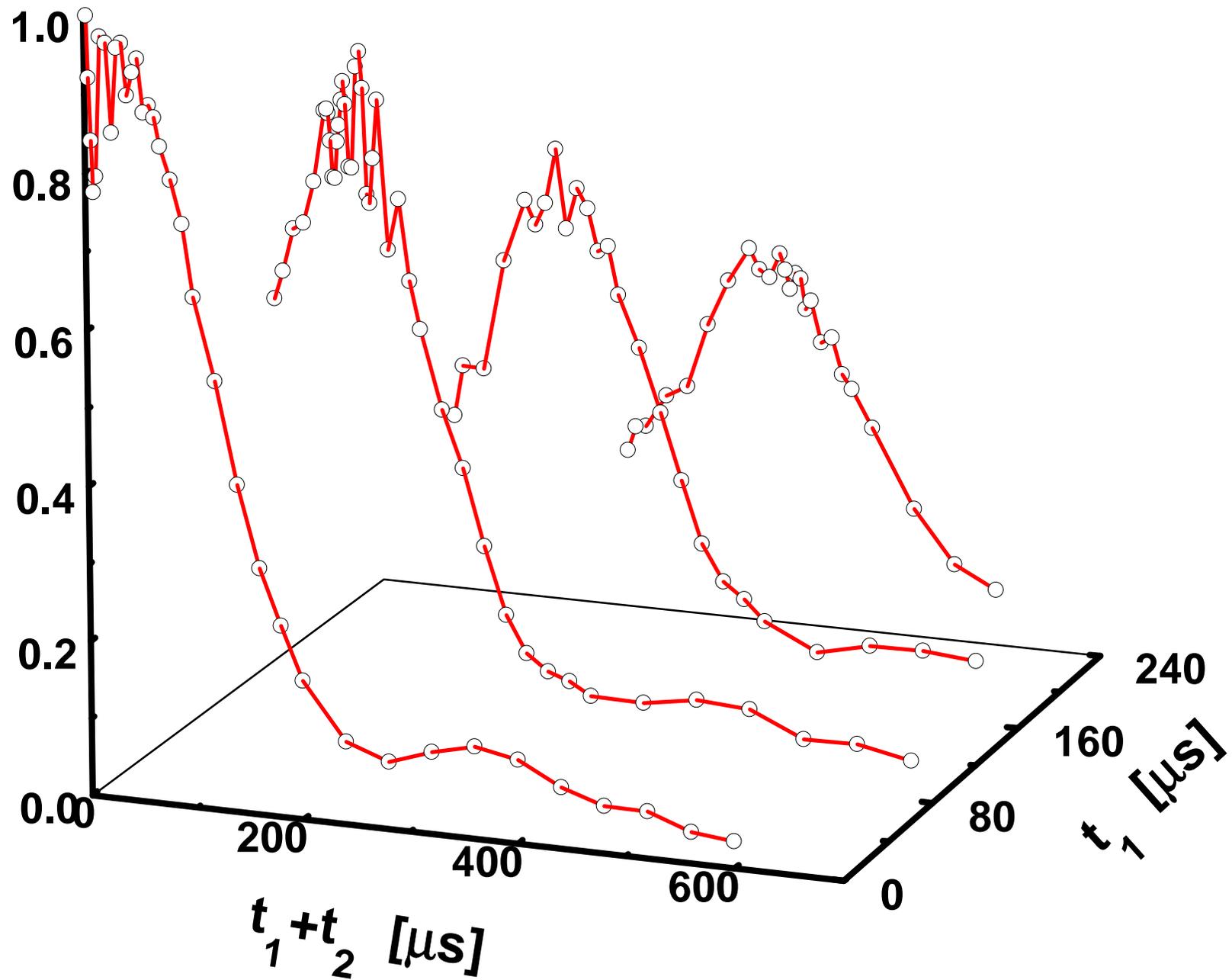
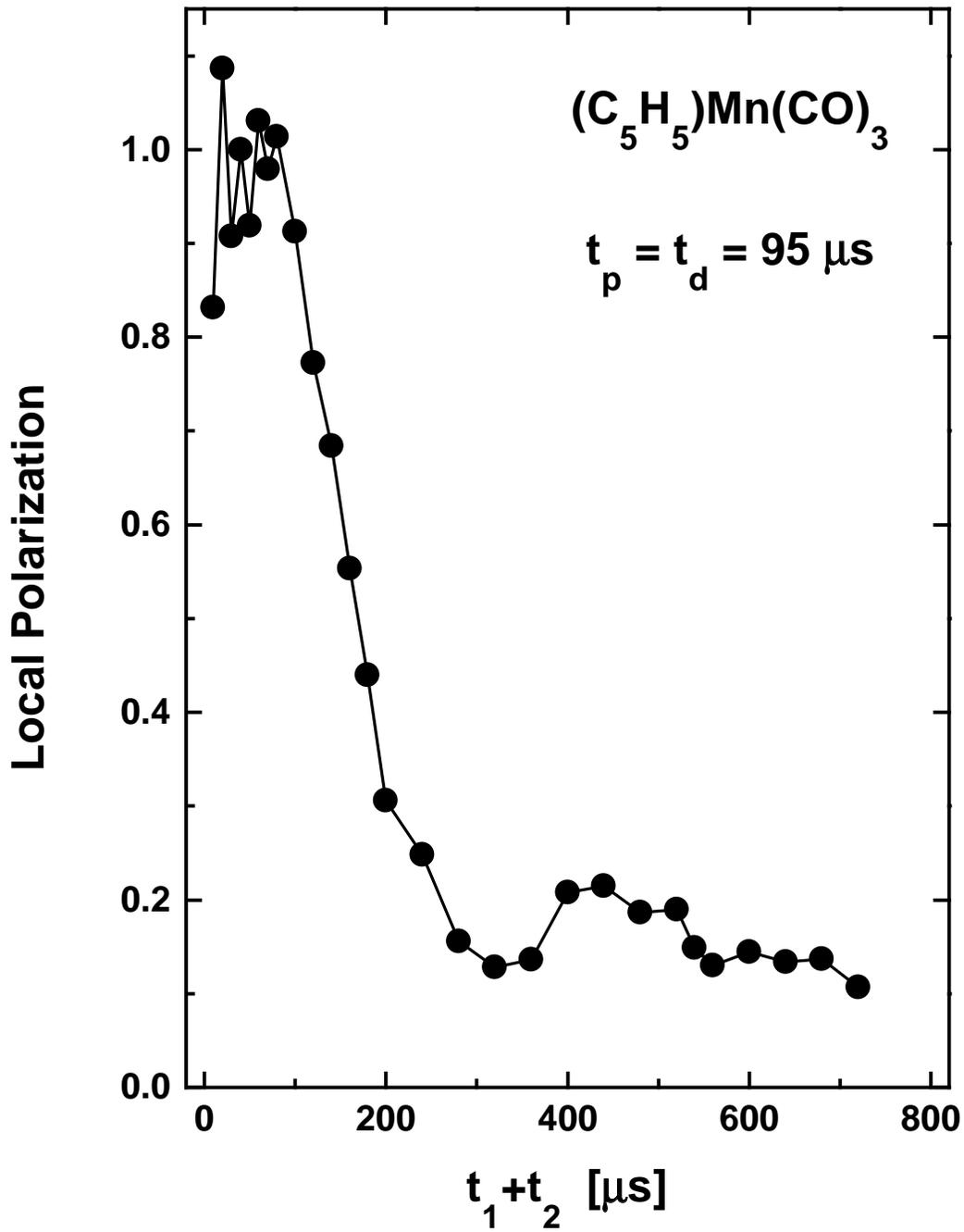
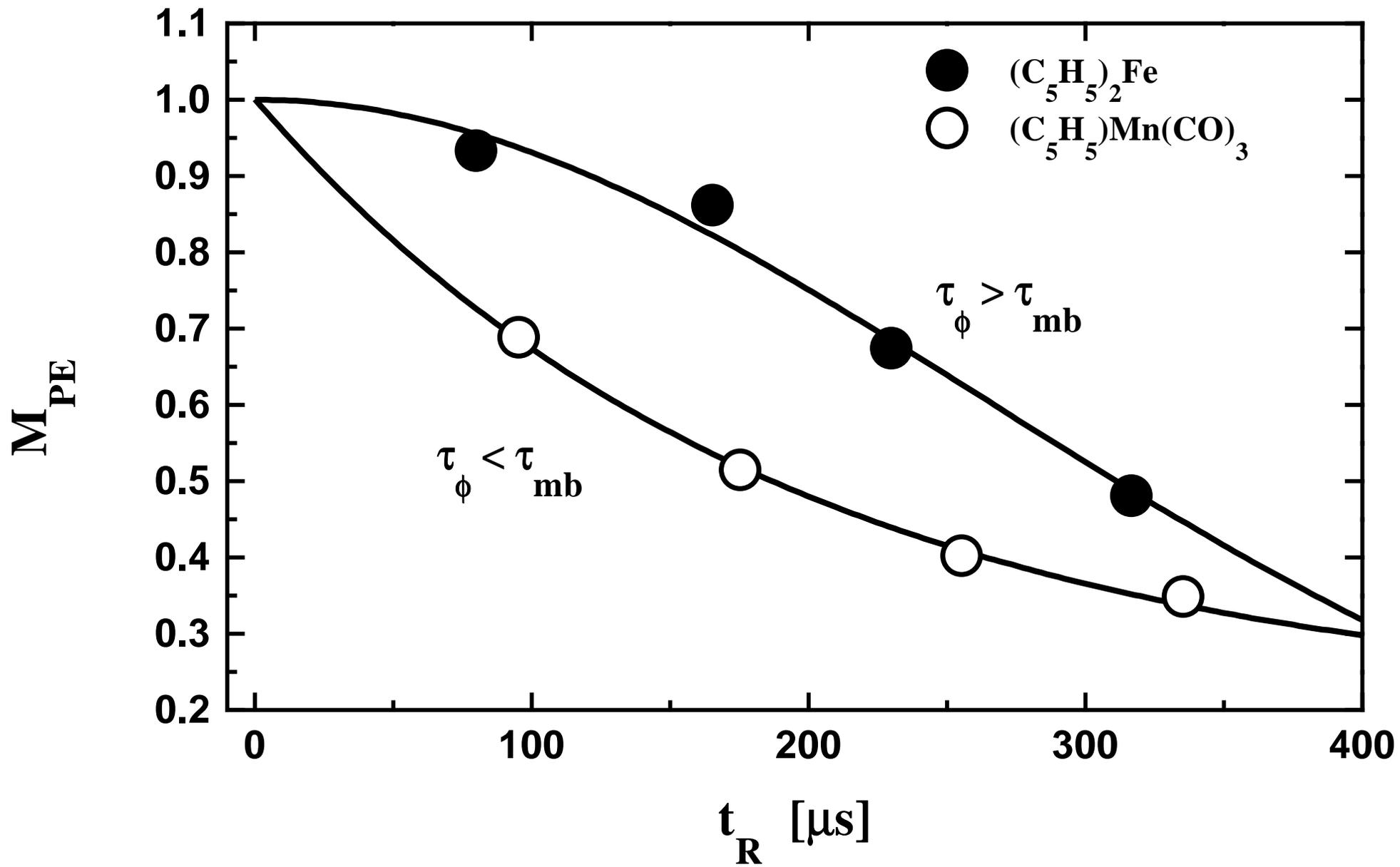
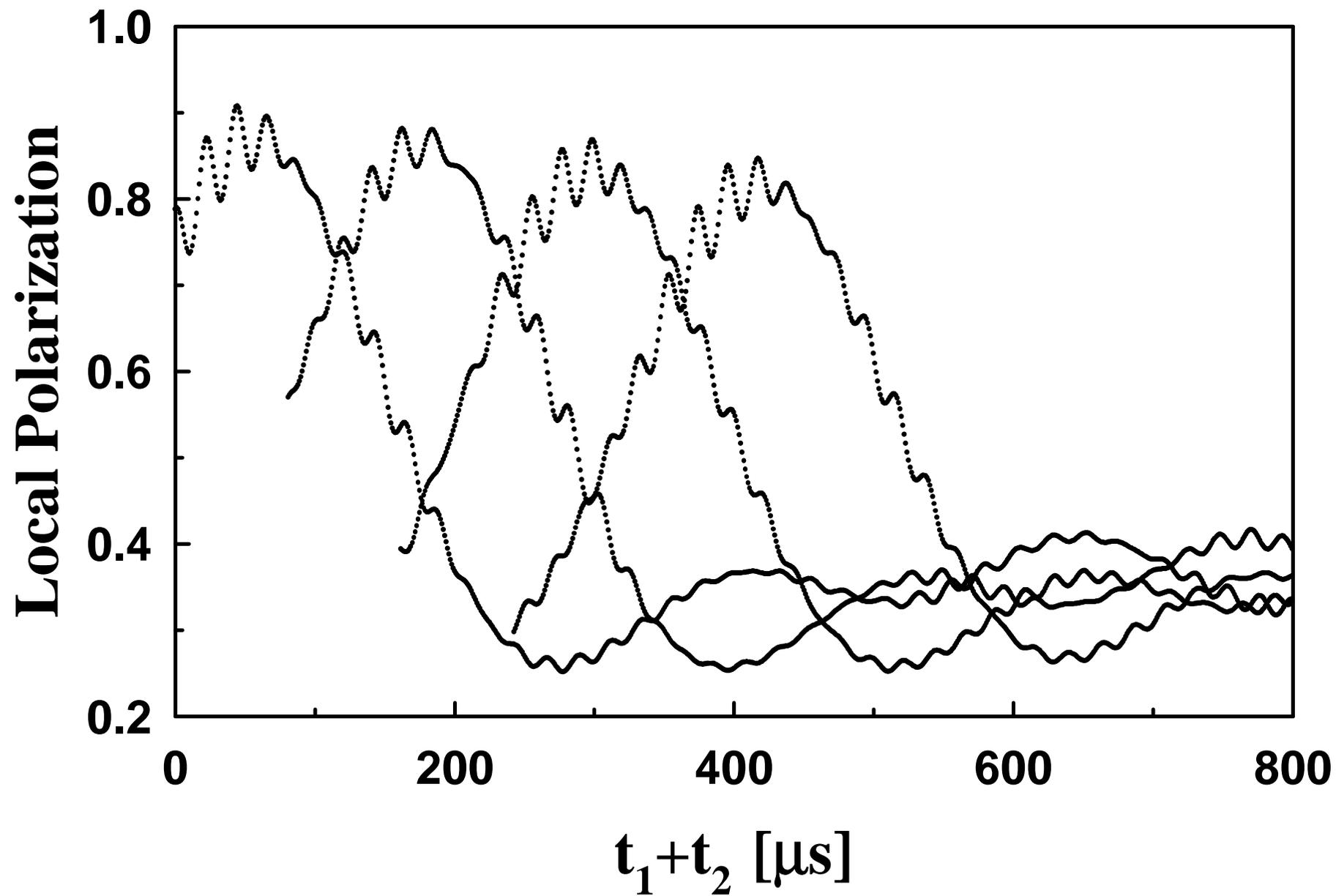
\end{document}